\documentclass[aps,prl,twocolumn]{revtex4}
\usepackage{graphicx}

\newcommand{\bra}[1]{\left\langle #1\right|}
\newcommand{\ket}[1]{\left| #1\right\rangle}
\newcommand{\braket}[2]{\left\langle #1|#2\right\rangle}
\newcommand{\ketbra}[2]{\left| #1\right\rangle\!\left\langle#2\right|}

\begin{document}                

\title{Remote preparation of a single-mode photonic qubit by measuring field quadrature noise}

\author{S. A. Babichev, B. Brezger, A. I. Lvovsky\cite{Lvovsky}}

\address{Fachbereich Physik, Universit\"at Konstanz, D-78457 Konstanz, Germany}

\date{\today}


\begin{abstract}

An electromagnetic field quadrature measurement, performed on one
of the modes of the nonlocal single-photon state
$\alpha\ket{1,0}-\beta\ket{0,1}$, collapses it into a
superposition of the single-photon and vacuum states in the other
mode. We use this effect to implement remote preparation of
arbitrary single-mode photonic qubits conditioned on observation
of a preselected quadrature value. The quantum efficiency of the
prepared qubit can be higher than that of the initial single
photon.

\end{abstract}

\maketitle

\label{sec:Introduction}


{\it Remote state preparation} (RSP) is a quantum communication
protocol which allows indirect transfer of quantum information
between two distant parties by means of a shared entangled
resource and a classical channel. Unlike the celebrated
teleportation scheme \cite{telep}, the sender (Alice) does not
possess a copy of the source state, but is aware of its full
classical description. To implement RSP, she performs a
measurement on her share of the entangled resource in a basis
chosen in accordance with the state she wishes to prepare.
Dependent on the result of her measurement, the entangled ensemble
collapses either onto the desired state at the receiver (Bob's)
location, or can be converted into it by a local unitary
operation.

Although RSP has been formulated  \cite{Lo-Bennett} and
investigated theoretically \cite{Pati-Paris-Bennett03} as a
quantum communication protocol only recently, its concept can be
traced back to the seminal work of Einstein, Podolsky, and Rosen
(EPR) \cite{EPR}, who have considered an entangled state of two
particles with correlated positions and momenta. By choosing to
measure either the position or the momentum of her particle, Alice
can remotely prepare Bob's particle in an eigenstate of either
observable, thus instantaneously creating either of two mutually
incompatible physical realities at a remote location.

Aside from many experiments on the EPR paradox, both in the
original \cite{EPRexp} and in the Bohm-Bell \cite{Bell}
configurations, controlled collapse of an entangled wavepacket has
been utilized experimentally to prepare a single photon by means
of conditional measurements on a photon pair generated via
parametric down-conversion. When a single-photon detector, located
in one of the emission channels, registers a photon, the entangled
pair state collapses into a single photon in a well-defined
spatiotemporal mode travelling along the other emission channel.
This technique was proposed in 1986 \cite{PhotonCond} and has
since been employed in many experiments.


In most theoretical and experimental work on controlled state
collapse, the observable measured by Alice coincided with the one
that defines the entanglement basis. Upon the measurement, the EPR
state will collapse into an eigenstate of this observable. If Bob
measures the same observable as Alice, his result will be highly
correlated with Alice's.

In the experiment reported here, such a straightforward
correlation is not present. We start with a two-mode optical state
entangled in the photon number (Fock) basis,
$\ket{\Psi}=\alpha\ket{1}_A\ket{0}_B-\beta\ket{0}_A\ket{1}_B)$. To
perform RSP, Alice measures the quantum noise of the electric
field quadrature observable $\hat X_{\theta A}$, which is only
weakly correlated in this ensemble. A particular result obtained
by Alice does not mean that Bob, by measuring the same observable,
would acquire the same (or similar) value. Yet, as we demonstrate,
the RSP scheme is fully functional: the measurement by Alice
collapses the EPR state into a pure single-mode photonic qubit
$x\ket{0}+y\ket{1}$.

\begin{figure}
    \begin{center}
        \includegraphics[width=0.45\textwidth]{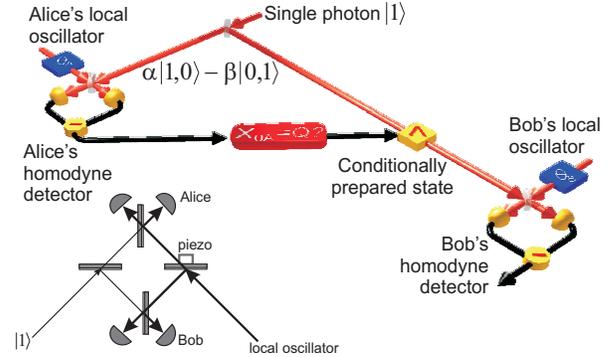}
        \caption{a) Scheme of the experiment. A photon
        incident on a beam splitter generates a nonlocal single-photon state. Alice performs a quadrature measurement
        using her homodyne detector, thus preparing a state $x
        \ket{0} + y \ket{1}$ at Bob's location. Inset: Schematic of the actual
        experimental arrangement. The four beam splitters must be
        kept interferometrically stable with respect to each other.}
        \label{Scheme}
    \end{center}
\end{figure}

In other words, we prepare a bit of quantum information encoded in
a {\it discrete} (photon number) basis by measuring a {\it
continuous} observable (field quadrature). So far, discrete- and
continuous-variable quantum information science has developed with
little overlap between these two domains. One of the main messages
of this Letter is that these two subfields are in fact closely
intertwined and that a number of novel phenomena can be observed
at their interface.

Other work investigating this interface include theoretical
proposals to improve the degree of squeezing in a two-mode
squeezed state \cite{SqDistill} and generating Schr\"odinger-cat
states \cite{cat}. Foster {\it et al.} performed a cavity QED
experiment in which detection of a photon coming out of a cavity
prepared an optical state with a well-defined phase \cite{Orozco}.

A {\it conceptual scheme} of our experiment is shown in Figure
\ref{Scheme}. A single photon $\ket{1}$ incident upon a beam
splitter with transmission $\alpha^2$ and reflection $\beta^2$
generates the entangled state $\ket{\Psi}$ which is shared between
Alice and Bob. With each incoming photon, Alice performs a
homodyne measurement on her part of the entangled state with the
local oscillator set to a preselected phase $\theta_A$. If her
measurement result is equal to a preselected value $Q$ (which we
call {\it conditional quadrature}), she notifies Bob via a
classical channel. Upon receipt of Alice's message, Bob performs a
homodyne measurement of his fraction of $\ket{\Psi}$ to
characterize the remotely prepared state.

Alice's homodyne measurement is associated with the quadrature
operator
\begin{equation} \hat X_{\theta A}=\hat X\cos
\theta_A+\hat P\sin \theta_A, \label{Xtheta} \end{equation} $\hat
X$ and $\hat P$ being the canonical position and momentum
observables. By detecting a particular quadrature value $X_{\theta
A}=Q$, Alice projects the entangled resource $\ket{\Psi}$ onto a
quadrature eigenstate $\bra{Q_{\theta A}}$:
\begin{eqnarray}
    \ket{\psi_B}&=&\mathcal{N}\braket{Q_{\theta A}}{\Psi}\\
    \nonumber&=&\mathcal{N}\left[\alpha \braket{Q_{\theta A}}{1}_A \ket{0}_B -\beta \braket{Q_{\theta A}}{0}_A
    \ket{1}_B\right],
\end{eqnarray}
which is just a coherent superposition of the single-photon and
vacuum states
\begin{eqnarray}\label{psiB}
    \ket{\psi_B} &=& x \ket{0} + y \ket{1}
\end{eqnarray}
with $x=\mathcal{N}\alpha \braket{Q_{\theta A}}{1}$ and
$y=-\mathcal{N}\beta\braket{Q_{\theta A}}{0}$ ($\mathcal{N}$ is a
normalization factor). These coefficients are the well known
stationary solutions of the Schr\"odinger equation for a particle
in a harmonic potential \cite{proof}:
\begin{eqnarray}
\braket{Q_{\theta}}{0}=\left(\frac{2}{\pi}\right)^{1/4}e^{-Q^2};\label{Q0}\\
\braket{Q_{\theta}}{1}=2\left(\frac{2}{\pi}\right)^{1/4}Qe^{-i
\theta}e^{-Q^2} .\label{Q1}
\end{eqnarray}
By choosing particular values of $\theta$ and $Q$, Alice can
remotely prepare any random vector on the surface of the Bloch
sphere.

In our {\it experiment} the initial single-photon state was
prepared by means of a conditional measurement on a biphoton
produced via parametric down-conversion. We used frequency-doubled
2-ps pulses from a mode-locked Ti:Sapphire laser running at
$\lambda=790$ nm which underwent down-conversion in a BBO crystal,
in a type-I frequency-degenerate, but spatially non-degenerate
configuration. A single-photon detector (Perkin-Elmer), placed
into one of the outcoming channels, detected biphoton creation
events and triggered the quantum communication protocol described
above. A more detailed description of our laser setup can be found
in Refs. \cite{FockMM,Shapiro}.

The actual geometric arrangement of the RSP apparatus is shown in
the inset to Fig.~1. The experiment required high interferometric
stability of all modes involved, so the distance between Alice's
and Bob's stations had to be minimized. Because the single-photon
state has no optical phase, only the relative phase between
Alice's and Bob's modes $\theta_A-\theta_B$ has a physical meaning
and affects the homodyne statistics. We have therefore chosen to
control this difference directly rather than each phase
individually. This was done by means of a piezoelectric transducer
as shown in the figure.

The local oscillator pulses for homodyne detection have been
provided by the master Ti:Sapphire laser. Their spatiotemporal
modes had to match the respective modes of the nonlocal
single-photon state. Mode matching was optimized via the technique
described in \cite{FockMM}, i.e. by simulating the single photon
by a classical pulse and maximizing the visibility of the
interference with the local oscillators at each beam splitter.

Optical losses, dark counts of the trigger detector, and non-ideal
mode matching result in some distortion of the RSP scheme.
Fortunately, almost all these imperfections can be accounted for
by assuming that the single photon entering the first beam
splitter has some admixture of the vacuum state:
\begin{equation}
  \hat{\rho}_{\ket{1}}=\ketbra{1}{1}+(1-\eta)\ketbra{0}{0}
\label{SINGLE}
\end{equation}
where $\eta_{\ket{1}}$ is the cumulative quantum efficiency
incorporating the imperfections of the entire apparatus. It can be
evaluated by independent reconstruction of optical ensembles
arriving to each homodyne detector. Both Alice's and Bob's
ensembles are statistical mixtures of the type (\ref{SINGLE}),
with the single photon fractions of $\alpha^2\eta$ and
$\beta^2\eta$, respectively. We found $\eta=0.55$ \cite{etanote}.
Note that Eq. (\ref{SINGLE}) is valid even though some of the
losses occur {\it after} the photon has been split into two modes.

Regarding inefficiencies according to Eq. (\ref{SINGLE}), we write
the remotely prepared ensemble as
\begin{eqnarray}\label{rhoB}
    \hat\rho_B=E\ketbra{\psi_B}{\psi_B}+(1-E)\ketbra{0}{0}
\end{eqnarray}
where $\ket{\psi_B}$ is given by Eq. (\ref{psiB}) and the qubit
preparation efficiency is
\begin{eqnarray}\label{E}
E=\frac{\eta(\alpha^2\braket{Q}{1}^2+\beta^2\braket{Q}{0}^2)}
{\eta(\alpha^2\braket{Q}{1}^2+\beta^2\braket{Q}{0}^2)+(1-\eta)\braket{Q}{0}^2}.
\end{eqnarray}
The ``success rate", i.e. the fraction of those events in which
$X_{\theta A}$ approximates $Q$, is proportional to
\begin{eqnarray}
R=(1-\eta)\braket{Q}{0}^2+\eta\alpha^2\braket{Q}{1}^2+\beta^2\braket{Q}{0}^2.
\end{eqnarray}

Our {\it data acquisition} procedure was based on postselection.
Homodyne measurements at Bob's station were performed every time,
independent of Alice's result. We varied the relative phase
$\theta_A-\theta_B$ slowly and with each incoming photon, acquired
a pair of values $(X_{\theta A},X_{\theta B})$ from both homodyne
detectors. Then we selected those pairs for which $X_{\theta A}$
approximated a particular conditional quadrature $Q$ within a
certain margin of error (Fig.~2) and reconstructed the optical
ensemble associated with the respective Bob's data.

For reconstruction, we utilized the novel likelihood-maximization
method \cite{MaxLik}. This technique, previously not applied to
experimental homodyne tomography, warrants a higher reconstruction
fidelity than the inverse Radon transformation employed
traditionally, and ensures physical plausibility of the
reconstructed ensemble. A detailed description of the
reconstruction procedure will be published elsewhere.

We have executed two data acquisition runs using two different
beam splitters with transmission $\alpha^2$ equal to 0.5 and 0.08.
With each beam splitter, a large data set of about 300,000 points
was acquired for a full relative phase cycle. The data were binned
up according to the value of $X_{\theta A}$ with the bin size of
0.071, except the last two bins which were twice as wide.
Maximum-likelihood estimation of Bob's ensembles associated with
each bin yielded density matrices in the Fock basis. As expected,
all matrix elements except $\rho_{00}$, $\rho_{01}$, $\rho_{10}$,
and $\rho_{11}$ were negligibly small. This allowed us to
interpret the reconstructed ensembles in accordance with Eq.
(\ref{rhoB}), i.e. as a statistical mixture of the state
$\ket{\psi_b}$ and the vacuum, and to evaluate the qubit value
$|y^2|$ and its preparation efficiency $E$ for each postselected
subset of the experimental data. The success rate is given by the
relative size of each subset. These quantities, along with their
theoretical predictions, are plotted in Fig.~3. Good agreement
between theory and experiment is achieved, except for relatively
high $Q$ values where the preparation rate is reduced and so are
the respective data subsets.

Fig.~3(a) shows that the fraction $|y^2|$ of the single photon in
the qubit decreases with increasing conditional quadrature $Q$.
This is easily interpreted by reviewing the vacuum and
single-photon wavefunctions (\ref{Q0}) and (\ref{Q1}). The
quadrature probability density associated with the single-photon
state is generally broader than that of the vacuum and vanishes at
$Q=0$. If Alice detects $X_A=0$, she can tell with certainty that
her mode is in the vacuum state and the photon must have been
reflected to Bob. On the contrary, detection of a large quadrature
value by Alice is much more likely if her mode contains a photon
--- and Bob's does not.

As evidenced by Fig.~3, a highly reflective ($\alpha^2=0.08$) beam
splitter provides a more profitable preparation rate and
efficiency for qubits with a high single-photon fraction (low $Q$)
than a symmetric beam splitter. For qubits with a high vacuum
fraction the relation is inverse. This is further illustrated in
Fig.~4 where experimental Wigner functions of two ensembles
prepared using different beam splitters and $Q$ values are
plotted.

\begin{figure}[tbp]
    \begin{center}
        \includegraphics[width=0.45\textwidth]{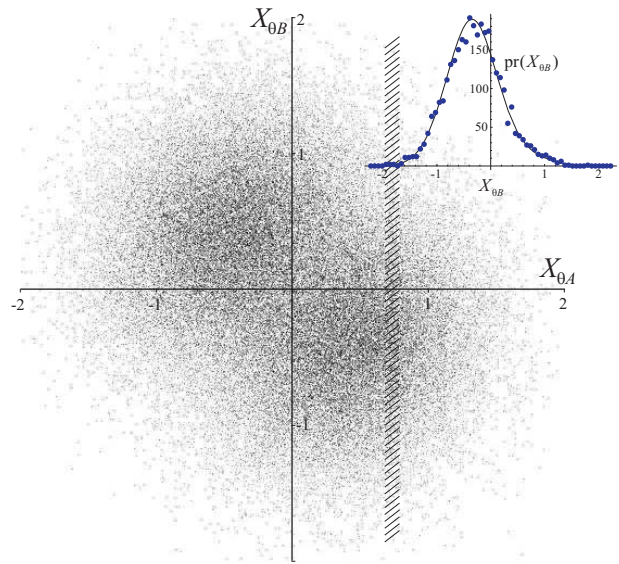}
        \caption{61440 samples of homodyne measurements by Alice and Bob acquired for
        $\theta_A-\theta_B=0$. Although both parties measure the
        same quadrature, the
        quantum noise exhibits little correlation. Inset: a
        histogram of Bob's data conditioned on $X_{\theta A}=0.71$
        (shaded area of the main plot). This is one of the marginal
        distributions of the Wigner function in Fig. 4(b).}
        \label{Nlproject}
    \end{center}
\end{figure}

\begin{figure}
    \begin{center}
        \includegraphics[width=0.37\textwidth]{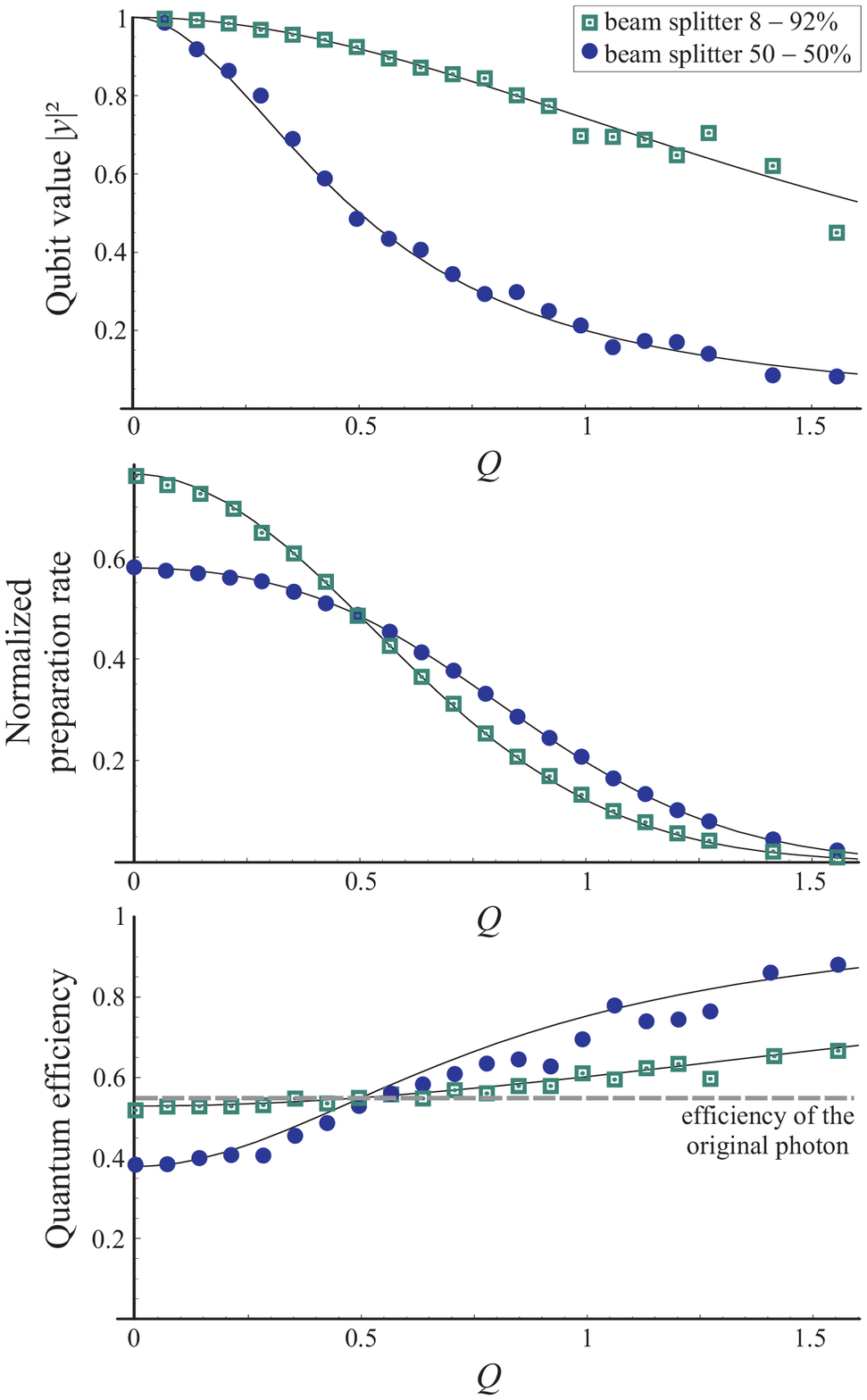}
        \caption{(a) Single-photon fraction $|y^2|$ in the qubit as a function of the
        preselected quadrature $Q$. With a
        symmetric beam splitter, values of $Q$ below $1/2$
        correspond to a prepared state with a single photon
        fraction greater than 50 percent. (b) Relative success rate of
        remote state preparation. (c) Quantum efficiency of the
        remotely prepared state. For $Q>1/2$ the efficiency of the output state is
        higher than that of the initial single photon.} \label{Qvalue}
    \end{center}
\end{figure}

\begin{figure}[tbp]
    \begin{center}
        \includegraphics[width=0.45\textwidth]{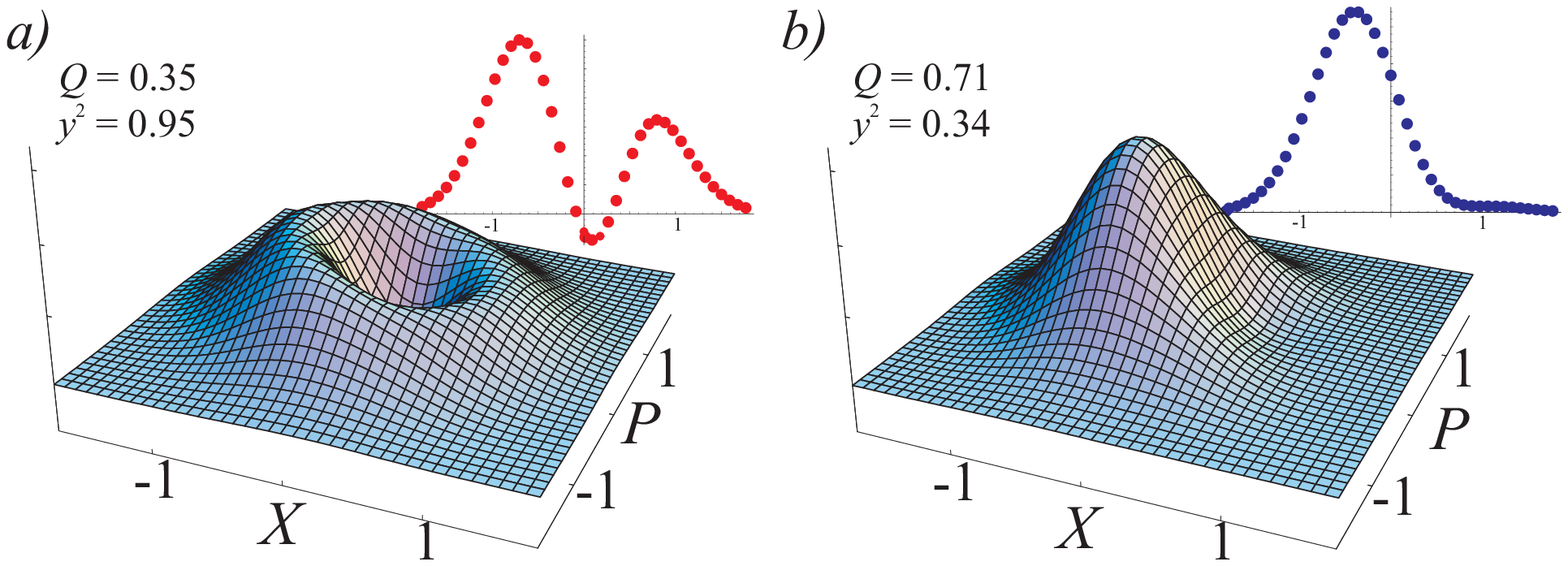}
    \end{center}
    \caption{Examples of Wigner functions of the remotely prepared
    ensembles reconstructed from the experimental data. (a): $Q=0.35$, highly reflective beam splitter; (b)
    $Q=0.71$, symmetric beam splitter. Insets show cross-sections
    through symmetry planes.} \label{Wigner}
\end{figure}

One surprising feature associated with the protocol is that the
preparation efficiency $E$ of the remotely prepared state can be
{\it higher} then the efficiency of the initial single photon as
long as the conditional quadrature value $Q$ exceeds $1/2$
 (Fig. 3(c)). In other words, the reported
scheme features not only preparation, but also, for qubits with a
sufficiently high vacuum fraction, {\it purification} of the
prepared qubit \cite{cirac-pur}. In quantum-optical experiments at
visible wavelengths the vacuum state is readily available; still
it appears surprising that this ``free" vacuum can be incorporated
into the prepared qubit in a controlled, coherent manner.

The observed purification effect raised our curiosity about a
possibility of extension to single photons. Can one distill a
high-purity single-photon state from a large set of mixtures
(\ref{SINGLE}) with moderate efficiency? This problem is relevant
to a variety of recently reported solid-state sources which are
capable of generating single photons ``on demand" but in a poor
spatiotemporal mode \cite{turnstyle}. Purification would make such
sources applicable to the linear optical quantum computation
scheme \cite{KLM}.

In conclusion, we have reported remote state preparation of
single-mode photonic quantum bits in a counterintuitive scheme. We
started with a two-mode quantum state with the entangled {\it
discrete} degree of freedom (number of photons), and by measuring
a {\it continuous} observable (field quadrature) in one of the
modes collapsed the entangled state into a coherent superposition
of two Fock states in the other mode, again in the {\it discrete}
domain. Surprisingly, the quantum efficiency of the prepared qubit
can be higher than that of the initial photon.

This experiment demonstrates, in our opinion, the potential of
combining discrete and continuous variable techniques in quantum
information technology applications.

We thank the Deutsche Forschungsgemeinschaft and the Optik-Zentrum
Konstanz for support.




\begin{references}
\bibitem{proof} Eqs. (\ref{Q0}) and (\ref{Q1}) for an arbitrary quadrature
(\ref{Xtheta}) follow from the properties of the phase-space
rotation operator
$\hat U(\theta)=e^{-i\theta \hat
n}$. From $\hat U^\dagger(\theta)\hat a\hat U (\theta)=\hat a
e^{-i\theta}$ we find for the quadrature operator $\hat
U^\dagger(\theta)\hat X\hat U (\theta)=\hat X_\theta$ and for its
eigenstate $\ket{Q_\theta}=\hat U^\dagger\ket{X}$. From the first
and last relations above, we obtain
$\braket{n}{Q_\theta}=e^{i\theta n}\braket{n}{X}$. The quantity
$\braket{n}{X}$ is the energy eigenwavefunction of a harmonic
oscillator.
\bibitem{etanote} A reduction from the value $\eta=0.62$ reported
in some of our earlier papers \cite{Shapiro} might be caused by a
replacement of our pulsed laser by one with a poorer spatial mode.
\end{references}
\end{document}